\documentclass[twocolumn,preprintnumbers,amsmath,amssymb,superscriptaddress]{revtex4}

\usepackage{graphicx}
\usepackage{color}
\usepackage{dcolumn}
\usepackage{bm}

\begin{document}

\author{William N. Plick} \email{billplick@gmail.com}
\affiliation{CNRS LTCI, Departement Informatique et Reseaux, Telecom ParisTech,
23 Avenue d'Italie, CS 51327, 75214 Paris CEDEX 13, France}

\author{Mario Krenn}
\affiliation{Quantum Optics, Quantum Nanophysics, Quantum Information,
University of Vienna, Boltzmanngasse 5, Vienna A-1090, Austria}
\affiliation{Institute for Quantum Optics and Quantum Information, Boltzmanngasse 3, Vienna A-1090, Austria}

\title{Physical meaning of the radial index of Laguerre-Gauss beams}

\date{\today}

\begin{abstract}
\noindent The Laguerre-Gauss modes are a class of fundamental and well-studied optical fields. These stable, shape-invariant photons \--- exhibiting circular-cylindrical symmetry \--- are familiar from laser optics, micro-mechanical manipulation, quantum optics, communication, and foundational studies in both classical optics and quantum physics. They are characterized, chiefly, by two modes numbers: the azimuthal index indicating the orbital angular momentum of the beam \--- which itself has spawned a burgeoning and vibrant sub-field \--- and the radial index, which up until recently, has largely been ignored. In this manuscript we develop a differential operator formalism for dealing with the radial modes in both the position and momentum representations, and \--- more importantly \--- give for the first time the meaning of this quantum number in terms of a well-defined physical parameter: the ``intrinsic hyperbolic momentum charge''.      
\end{abstract}

\maketitle

\section{Introduction}

The ability to shape optical beams has increased dramatically in recent years thanks to a number of newly developed tools and techniques. This progress has been driven by the many promising potential applications of the orbital angular momentum degree of freedom, as well as more generalized beam shaping. In 1992 Allen et al. \cite{allen} showed the connection between the azimuthal index of a Laguerre-Gauss (LG) mode (the natural beam-like light modes exhibiting circular-cylindrical symmetry) and the physical quantity of orbital angular momentum (OAM). This physical parameter is useful for doing micro-mechanical work, is a well-defined quantum number which applications to foundations, and informatics, and also is important in the foundational study of optics (see Ref.\cite{rev} and references therein).

However, despite the vast amount of attention OAM has received over the past two decades very little research has been conducted on the \emph{other} mode number of LG photons \--- the radial index. Much of the time it is regarded as little more than unwanted noise which unfortunately arises when one is trying to produce ``pure'' OAM modes. In a previous note we referred to this as the ``forgotten quantum number'' \cite{plick}. In this paper we will summarize the previous work by us and by others \cite{k1,k2}, then significantly extend and simplify these previous analyses, elucidate the physical ``meaning'' of this quantum number, and give a brief prospectus for the use of the radial number in future quantum technologies.  

Laguerre-Gauss beams are typically found by taking the paraxial wave equation and finding solutions in the circular-cylindrical coordinate system. The paraxial equation is the result of making the assumption that the beam is not highly divergent or focused. Ironically, we find the \emph{mathematical} analysis of the radial number simplifies significantly when this assumption is removed and exact solutions of Maxwell's equations are considered. This is done by taking the photonic wave-function in momentum coordinates as derived through the Riemann-Silberstein vector formalism \cite{BB1,BB2}. However, on the other hand, the physical meaning does remain clearer in the position representation, an oddity we will discuss.

Our conclusion is that the radial index of Laguerre-Gauss photons is a compound physical parameter influenced by \--- but not influencing \--- other fundamental parameters of the mode, as well as an additional property intrinsic to itself: the hyperbolic momentum; which is a kind of mathematically well-formed ``radial-like'' momentum with subtle and interesting properties. Thus we call the radial index the ``hyperbolic momentum charge''. The hyperbolic momentum is the result of the restriction that the radial coordinate is only defined for values greater than zero, as we will explain.

This paper is organized as follows: In the following section we review the LG beams, in section 3 we derive the radial mode operator in the paraxial, position-space representation and introduce the hyperbolic momentum operator. In section 4 we briefly summarize the background for the exact quantum momentum-space wave-function of the photon, namely, the Riemann-Silberstein vector \--- a more detailed derivation being included in the appendix. In section 5 we derive the momentum-space formalism for the radial modes. In section 6 is the raison d'\^{e}tre of this manuscript: the physical interpretation of the preceding mathematics as well as a discussion of some potential applications of the radial quantum number. In section 7 we summarize and conclude.                

\section{Laguerre-Gauss beams and the orbital angular momentum of light}

The equation of a LG beam, under the paraxial assumption, in circular-cylindrical coordinates ($r$, $\phi$, $z$) is,

\begin{flalign}
&\mathrm{LG}_{nl}(r,\phi ,z)&&\label{LG}\\
&\quad =\sqrt{\frac{2n!}{\pi (n+|l|)!}}\frac{1}{w_{z}}\left(\frac{\sqrt{2}r}{w_{z}}\right)^{|l|}L_{n}^{|l|}\left(\frac{2r^{2}}{w_{z}^{2}}\right)&&\nonumber\\
&\quad\quad\times\mathrm{exp}\left[-\frac{r^{2}}{w_{z}^{2}}+i\left(l\phi+\frac{kr^{2}}{2R_{z}}-(2n+|l|+1)\varphi_{g}\right)\right].&&\nonumber
\end{flalign}

\noindent Where $n$ and $l$ are the radial and orbital angular momentum quantum numbers, respectively; $L_{n}^{|l|}$ is the generalized Laguerre polynomial of order $n$ and degree $|l|$. The functions $w_{z}$, $R_{z}$, and $\varphi_{g}$ are the beam waist, radius of curvature, and Gouy phase of the fundamental beam, and are given by

\begin{eqnarray}
w_{z}&=&w_{o}\sqrt{1+\frac{4z^{2}}{k^{2}w_{o}^{4}}},\\
R_{z}&=&z+\frac{k^{2}w_{o}^{4}}{4z},\\
\varphi_{g}&=&\mathrm{arctan}\left[\frac{2z}{kw_{o}^{2}}\right].\label{gouy}
\end{eqnarray}

\begin{figure*}[ht]
\includegraphics[scale=0.53]{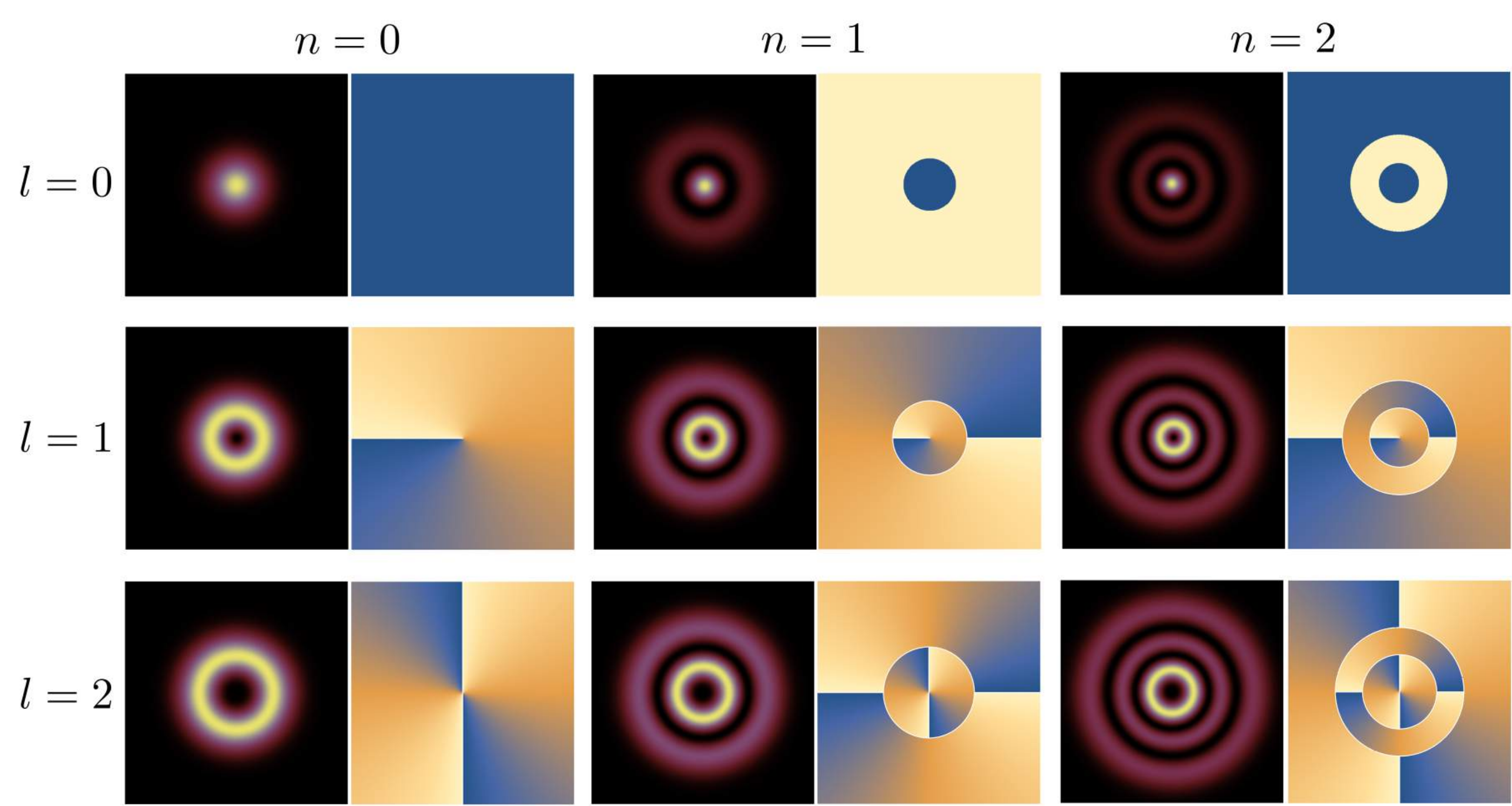}
\caption{The transverse spatial profiles of nine different Laguerre-Gauss beams in both intensity (left) and phase (right) at $z=0$, for three different values of the radial index and three different values of the azimuthal index. The diameter of all the plots is 6mm.}\label{lg}
\end{figure*}

\noindent Where $k=2\pi/\lambda$ is the overall wavenumber of the beam, and $w_{o}$ is the beam waist at $z=0$ (defined where the beam is narrowest). The beam or photon is then completely defined by four numbers $n$, $l$, $k$, $w_{o}$. In Figure \ref{lg} we plot some LG beams for various parameters. The LG beams are solutions to the paraxial wave equation (PWE)

\begin{eqnarray}
\nabla^{2}_{t}E-2ik\frac{\partial}{\partial z}E=0.\label{PWE}
\end{eqnarray}

\noindent Where $\nabla^{2}_{t}$ is the transverse Laplacian and $E$ is a complex electric scalar field (i.e. we assume the electric field vector points in the same direction at every point in the transverse plane). The paraxial wave equation is a version of the full wave-like Maxwell's equations under the small angle approximation for propagation. In this paper we will consider both the cases where this approximations is met, and when it may not be.

In the paraxial limit the angular momentum of light separates out into a spin and orbital part, both of which are well-defined \cite{calvo}. The form of the differential OAM operator in the paraxial limit, about the direction of beam-propagation, is well-known and straightforward

\begin{eqnarray}
\hat{L}_{z}=-i\hbar\frac{\partial}{\partial\phi}.\label{OAM}
\end{eqnarray}

\noindent The OAM is also sometimes called the ``winding number'' as it increments the number of helical equal-phase surfaces of the photon (see Fig.\ref{lg}). If a particle absorbs a photon with both spin (arising from the polarization of light) and orbital angular momentum the spin will cause the particle to rotate about it's own axis, whereas the OAM will cause it to rotate about the optical axis of the beam \cite{abs}. 

\section{The radial modes - operator formalism in the paraxial regime}

The orbital angular momentum of Laguerre-Gauss beams has received an extensive amount of attention, however the radial index ($n$ in our notation) has been the subject of only a handful of manuscripts. The radial index is so-called because the intensity pattern of LG beams exhibit $n+1$ concentric rings if $l\neq 0$ (there are $n$ rings around a central dot for $l=0$). Also, the phase structure of the beam displays $n$ concentric radial discontinuities with no smooth transitions at $z=0$ (see Fig.\ref{lg}). This is in contrast to the OAM/azimuthal coordinate where the phase does have smooth transitions at $z=0$. Unlike the OAM operator, Eq.\ref{OAM}, the differential operator for the radial number has not been studied until very recently.

The operator formalism for the radial modes of LG beams was first derived by Karimi and Santamato \cite{k1} in 2012 via group-theoretic techniques, and then independently by Plick et al. in 2013 \cite{plick} via the method described below. 

We start with the Laguerre polynomial. There exist a series of relations between Laguerre polynomials of varying order and degree (the so-called ``three-point relations''). One such relation is

\begin{eqnarray}
nL_{n}^{l}(x)=(l+1-x)L_{n-1}^{l+1}(x)-xL_{n-2}^{l+2}(x).
\end{eqnarray}

\noindent Which, when combined with the rule for differentiation of the polynomials,

\begin{eqnarray}
\frac{\partial}{\partial x}L_{n}^{l}(x)=-L_{n-1}^{l+1}(x),
\end{eqnarray}

\noindent yields,

\begin{eqnarray}
\left[(x-l-1)\frac{\partial}{\partial x}-x\frac{\partial^{2}}{\partial x^{2}}\right]L_{n}^{l}(x)=nL_{n}^{l}(x).\label{L}
\end{eqnarray}

\noindent Given this differential relation, it is possible to arrive at a relationship between the full LG modes by left-multiplying the other factors (the non-polynomial terms) in the LG function Eq.(\ref{LG}) onto Eq.(\ref{L}), and commuting those factors past the differentials on the left. Doing this we obtain

\begin{flalign}
&\left[-\frac{1}{4x}\frac{\partial^{2}}{\partial\phi^{2}}-x\frac{\partial^{2}}{\partial x^{2}}+\frac{i}{2}\frac{\partial}{\partial\phi}-\frac{\partial}{\partial x}+\frac{x}{4}-\frac{1}{2}\right]&&\\
&\quad\quad\quad\quad\quad\quad\quad\quad\quad\quad\quad\times\mathrm{LG}_{nl}(x,\phi)=n\mathrm{LG}_{nl}(x,\phi).&&\nonumber
\end{flalign}

\noindent Now if we make the the identification $x=2r^{2}/w_{o}^{2}$ and perform a change of coordinates, we obtain

\begin{flalign}
&\hat{N}_{o}\equiv\left[-\frac{w_{o}^{2}}{8}\left(\frac{1}{r^{2}}\frac{\partial^{2}}{\partial\phi^{2}}+\frac{1}{r}\frac{\partial}{\partial r}+\frac{\partial^{2}}{\partial r^{2}}\right)\right.&&\label{comp}\\
&\quad\quad\quad\quad\quad\quad\quad\quad\quad\quad\quad\quad\left.+\frac{i}{2}\frac{\partial}{\partial\phi}+\frac{1}{2}\left(\frac{r^{2}}{w_{o}^{2}}-1\right)\right].\nonumber&&
\end{flalign}
  
\noindent Where, $\hat{N}_{o}\mathrm{LG}_{nl}(r,\phi,0)=n\mathrm{LG}_{nl}(r,\phi,0)$, and we have defined the differential $n$-mode operator for $z=0$. A trivial transformation shows that this is indeed the operator derived in \cite{k1}.

Further simplification is possible by identification with other, better known, operators yielding the opportunity for some physical insight

\begin{eqnarray}
\hat{N}_{o}=-\frac{w_{o}^{2}}{8}\nabla^{2}_{t}-\frac{\hat{L}_{z}}{2}+\frac{1}{2}\left(\frac{r^{2}}{w_{o}^{2}}-1\right).\label{N0}
\end{eqnarray}

\noindent The operator is composed of the transverse Laplacian in cylindrical coordinates, and thus has a direct relationship to the transverse momentum squared (up to some constants). For completeness this Laplacian is defined as 

\begin{eqnarray}
\nabla^{2}_{t}=\frac{1}{r}\frac{\partial}{\partial r}r\frac{\partial}{\partial r}+\frac{1}{r^{2}}\frac{\partial^{2}}{\partial\phi^{2}}.\label{lap}
\end{eqnarray}

\noindent Equation (\ref{N0}) also contains the OAM operator (with $\hbar=1$), as well as a term reminiscent of a harmonic potential. It commutes with the OAM operator, but not with the Hamiltonian for free propagation due to the final potential-like term. This indicates that $n$ is \emph{not} a conserved quantity of the beam with respect to evolution through free space. This point will be discussed further in section 6.

It is also useful to derive a general operator for any value of $z$. Continuing with the same methods as above we have

\begin{eqnarray}
\hat{N}_{z}=-\frac{w_{z}^{2}}{8}\nabla^{2}_{t}+\frac{iz}{kw_{o}^{2}}\frac{\partial}{\partial r}r-\frac{\hat{L}_{z}}{2}+\frac{1}{2}\left(\frac{r^{2}}{w_{o}^{2}}-1\right).\label{zdep}
\end{eqnarray}

\noindent The differences being that the pre-factor of Laplacian is now the $z$-dependent beam waist and an additional term resultant from the phase imparted from the radius of curvature, $R_{z}$ (this can be seen from the fact that if the LG beam is written without the radius of curvature phase factor then this term does not appear). Interestingly, the final term is unchanged and remains $z$-independent.

The $z$-dependent form in Eq.(\ref{zdep}) offers another insight. The second term (radius-of-curvature term) can be identified as the operator for a well-formed radial momentum known as the ``hyperbolic momentum''. 

It has been known for a long time \cite{first} that the operator for a strictly radial momentum can not be well defined. To see this consider the most direct attempt: $\hat{p}_{r}=-i\hbar\partial_{r}$, and its action on the radial coordinate operator: 

\begin{eqnarray}
e^{i\gamma\hat{p}_{r}/\hbar}\hat{r}e^{-i\gamma\hat{p}_{r}/\hbar}=\hat{r}+\gamma.
\end{eqnarray}  

\noindent But the domain of $\hat{r}$ is only the non-negative real numbers. The operator $\hat{p}_{r}$ can take $\hat{r}$ out of this domain and is thus not well-formed. An operator which \emph{can} be well-defined in the circular-cylindrical coordinate system is the hyperbolic momentum

\begin{eqnarray}
\widehat{P_{H}}=-i\hbar\left(r\frac{\partial}{\partial r}+1\right).\label{hype}
\end{eqnarray}    

\noindent Which, up to constants (and the $z$ coordinate) is the second term in Eq.(\ref{zdep}). The action of this on the radial coordinate is 

\begin{eqnarray}
e^{i\gamma\widehat{P_{H}}/\hbar}\hat{r}e^{-i\gamma\widehat{P_{H}}/\hbar}=\hat{r}e^{\gamma}.
\end{eqnarray}  

\noindent Thus the hyperbolic momentum generates dilations, not linear translations, and can not cause the radial coordinate to be negative no matter the value of $\gamma$. As linear momentum is associated with invariance under translation, hyperbolic momentum is associated with invariance under scale transformations. However, since $\left[\nabla^{2}_{t},\widehat{P_{H}}\right]=-2i\hbar\nabla^{2}_{t}$ hyperbolic momentum is not a conserved property of paraxial photon propagation (unlike, obviously, the linear-$z$ momentum and the OAM).  It is clear that in our case the position $r$ can not be negative, that is $r\in\Re^{+}$. A convenient method of dealing with this is to make the transformation $\eta=\mathrm{ln}(r)$, which is only defined for positive positions. The conjugate variable to $\eta$ is the hyperbolic momentum. With this construction it is unsurprising that the hyperbolic momentum should appear in our formalism. The hyperbolic momentum has many other interesting properties, for a detailed investigation the interested reader is referred to Ref.\cite{T1} and Ref.\cite{T2}.

\begin{figure}
\includegraphics[scale=0.41]{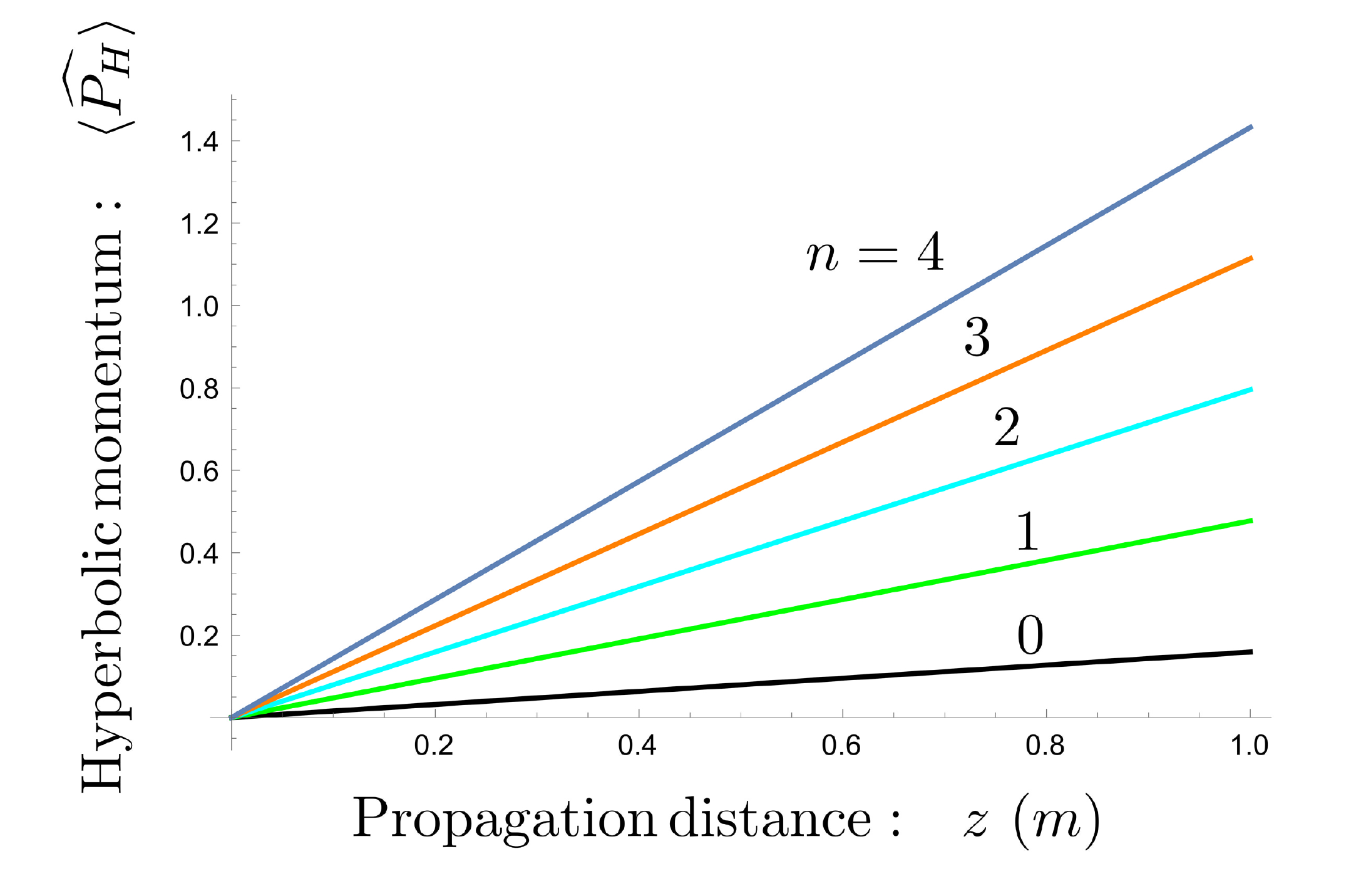}
\caption{The expectation value of the hyperbolic momentum as a function of propagation distance for five different values of the radial index. In all cases the scaling is linear, passing through the origin where the beam-waist is smallest. Note that this does not include the pre-factor of $-z/kw_{o}^{2}$ in Eq.(\ref{zdep}). For that whole term together the $z$-scaling is quadratic. Which, in the far field is equivalent to the scaling of the Laplacian term.}\label{phz}
\end{figure}

\begin{figure}
\includegraphics[scale=0.41]{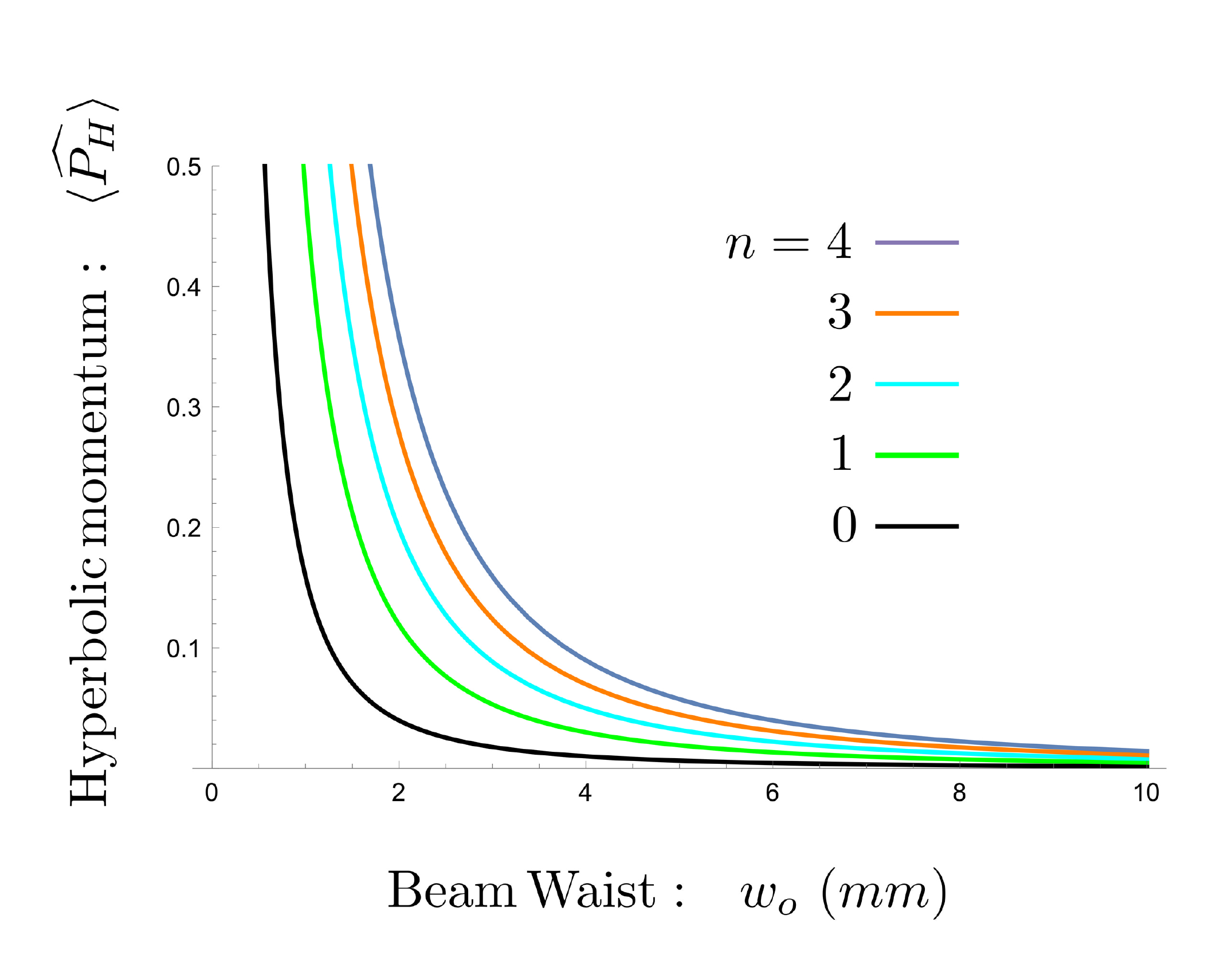}
\caption{The expectation value of the hyperbolic momentum as a function of beam-waist at origin for five different values of the radial index. The hyperbolic momentum has a finite value at $w_{o}$ (not pictured due to scaling in order to show contrast between modes).}\label{phw}
\end{figure}

For the case of a LG-beam the expectation value of the hyperbolic momentum increases linearly as a function of the propagation distance $z$. Thus the expectation value of the second term in Eq.\ref{zdep} as a whole has a quadratic scaling in $z$. The hyperbolic momentum is always zero where the beam-waist is narrowest, since at this point the beam is neither dilating nor contracting (see Fig.\ref{phz}). If we look at the same quantity as a function of beam waist $w_{o}$ we see an exponential decay. This later effect can be understood from the fact that as beam waist increases so does the degree of collimation of the beam, thus the beam dilates less (see Fig.\ref{phw}). These functions can be directly calculated as

\begin{eqnarray}
\langle\widehat{P_{H}}\rangle=\int d\vec{x}\,\mathrm{LG}^{*}_{nl}(r,\phi,z)\,\widehat{P_{H}}\,\mathrm{LG}_{nl}(r,\phi,z)\label{ph}
\end{eqnarray}

See Figures \ref{phz} and \ref{phw} for an illustration of the numerical results calculated from Eq.(\ref{ph}).

We will examine in more detail the physical interpretation of the operator in Eq.(\ref{zdep}) in section 6.  

It has been previously noted by Karimi and Santamato in Ref.\cite{k1} that the OAM and the radial index are inextricably linked. From the referenced paper: ``OAM and radial intensity distribution are strictly correlated, and different OAM generators produce specific (and different) distributions of radial modes.'' (We will comment again on this fact in section 6). In that paper ``radial coherent'' (displaced vacuum) and ``intelligent'' (minimum uncertainty) beams are also derived. The latter of which can be generalized to squeezed states which is done by Karimi et al. in Ref.\cite{k2} via a sophisticated algebraic technique employing the raising and lowering operators on the radial number \--- which form a lie algebra. It is also possible to alternatively define the coherent state as the eigenstate of the lowering operator, which is also done in Ref.\cite{k1}. Unlike light in the number basis \--- where all three definitions are connected via the familiar coherent state $|\alpha\rangle$ and its generalization the single-mode squeezed state \--- in the radial representation, all three concepts result in distinct beams.  

There has also been experimental work done on the value of the radial index as a quantum number. Again this was carried out by Karimi and collaborators in Ref.\cite{ke}. They showed via Hong-Ou-Mandel interference that the radial degree of freedom is indeed a quantum number and could in principle be used in quantum-information tasks. However, a small caveat is that care must be taken to ensure that the beam waists are equivalent. From that paper: ``The chosen basis is beam-waist dependent; an eigenstate for a specific beam waist turns into a superposition of radial modes for any other beam waist.'' 

Several experiments have taken advantage of both the OAM and radial indices, for example Ref.\cite{lof}. It has also been shown \cite{mario} that via the use of both quantum numbers, entangled states of very high dimension may be produced. Even as high as 103 dimensions. Research has also been done on efficiently producing radial modes \cite{walsh}. It has also recently been shown that LG beams of high radial indices exhibit self-healing properties in the same way that Bessel-beams do \cite{LGBB}.      

\section{The Riemann-Silberstein vector, exact solutions to Maxwell's equations, and the photonic wave-function}

Here we briefly outline some previously existing mathematical formalism we will need. A more complete version is included in an appendix. The Riemann-Silberstein (RS) vector is a representation of the electromagnetic field given by

\begin{eqnarray}
\vec{F}=\sqrt{\frac{\epsilon_{0}}{2}}\left(\vec{E}+ic\vec{B}\right).
\end{eqnarray}

\noindent Where $c$ is the speed of light in vacuum, $\epsilon_{0}$ is the permittivity of vacuum and $\vec{E}$ and $\vec{B}$ are the electric and magnetic fields, respectively. The vector consists of three complex numbers and many useful quantities can be calculated directly with simple equations. For a full review on the RS vectors the interested reader is directed to Ref.\cite{BB1} and Ref.\cite{BB2}. In the appendix we review this derivation in the interest of completeness and notational consistency. The previously acquainted reader may skip it.     

It is convenient to work with the scalar function $\chi$ (which contains the full information of $\vec{F}$, up to choice of gauge, as explained in the appendix) as opposed to the full RS vector $\vec{F}$. Given the formalism from the appendix we can directly write a general beam in the Bessel-Gauss basis as

\begin{eqnarray}
\chi^{\sigma}_{m}(r,\phi,z,t)=\int d\vec{k}\,\psi^{\sigma}(\,\vec{k}\,)e^{-i\sigma(ck-k_{z}z)}J_{m}(k_{t}r).
\end{eqnarray}

\noindent Using Eqs.(\ref{xmo},\ref{xm}). Note that we take a different definition than is used in Ref.\cite{beam} as we pull the azimuthal phase and jacobian into $\psi$. 

Now, derivation of the Laguerre-Gauss modes is facilitated by a change of momentum coordinates $k_{\pm}=(k\pm k_{z})/2$ and also time $t_{\pm}=t\pm z/c$, yielding

\begin{eqnarray}
\chi^{\sigma}_{m}(r,\phi,z,t)&=&\int_{0}^{\infty}dk_{+}\int_{0}^{\infty}dk_{-}\int^{2\pi}_{0}dk_{\phi}\nonumber\\
& &\times \psi^{\sigma}(k_{+},k_{-}) e^{-i\sigma c(k_{+}t_{-}+k_{-}t_{+})}\nonumber\\
& &\times J_{m}(2r\sqrt{k_{+}k_{-}}).\label{pos}
\end{eqnarray}  

\noindent It is of key importance to note that since $\chi$ is effectively the position-space wave-function (given the caveats described in the appendix) that $\psi^{\sigma}(k_{+},k_{-})$ is the \emph{momentum-space wave-function} in the beam-like (i.e. Bessel) basis. Now, if we make the choice

\begin{eqnarray}
\psi^{\sigma}(k_{+},k_{-},k_{\phi})&=&e^{i\sigma mk_{\phi}}\delta\left(k_{+}-\frac{\Omega}{c}\right)\nonumber\\
& &\times k_{-}^{n+\frac{|m|}{2}}e^{-\frac{w^{2}\Omega}{c}k_{-}}k.\label{momwave}
\end{eqnarray}   

\noindent Where $n$ is some integer, and $w$ is some real number, we can solve Eq.(\ref{pos}) exactly:

\begin{eqnarray}
\chi^{\sigma}_{m}(r,\phi,z,t)&=&\frac{\mathcal{N}\,r^{|m|}}{a(t_{+})^{n+|m|+1}}e^{-i\sigma\Omega(t-z/c-m\phi)}\nonumber\\
& &\times e^{-\frac{r^{2}}{a(t_{+})}}L^{|m|}_{n}\left(\frac{r^{2}}{a(t_{+})}\right).
\end{eqnarray}

\noindent Which is clearly a form of the Laguerre-Gauss beams, with $\mathcal{N}$ a normalization constant, $m$ an angular momentum number (we avoid calling this the orbital angular momentum since the OAM is only completely well-defined in the paraxial limit), $n$ the radial number, and $a(t_{+})=w^{2}+i\sigma c^{2}t_{+}/\Omega$ \--- a form of the complex beam parameter. Thus we can identify Eq.(\ref{momwave}) as the exact momentum-space wave-function of a Laguerre-Gauss beam (up to normalization constants). 

\section{The momentum-space formalism for the radial modes}

In the previous section, and in the appendix, we summarized the mathematical tools from Ref.\cite{beam} which we will need for the investigation of the radial number as it exists for the exact photon wave-function. 

From the momentum representation of Laguerre-Gauss beams in Eq.(\ref{momwave}), we can easily derive an radial momentum operator.

\begin{align}      
k_- \frac{\partial}{\partial k_-} \psi^{\sigma} = \Big(\frac{k_-}{k} + n + \frac{m}{2} -\frac{w^{2}\Omega}{c} k_- \Big) \psi^{\sigma},\\
\Big(k_- \frac{\partial}{\partial k_-} + \frac{i}{2\sigma}\frac{\partial}{\partial k_{\phi}} - \frac{k_-}{k} +\frac{w^{2}\Omega}{c}  k_- \Big) \psi^{\sigma} = n \psi^{\sigma}.   
\end{align}

\noindent Therefore, one finds the (general, non-paraxial) radial momentum operator as 

\begin{align}  
\hat{N}_{k} = \Big(k_- \frac{\partial}{\partial k_-} + \frac{i}{2\sigma}\frac{\partial}{\partial k_{\phi}} - \frac{k_-}{k} +\frac{w^{2}\Omega}{c}  k_- \Big).
\end{align}
    
\noindent In order to interpret the operator, one can transform it to polar coordinates. By using the relations

\begin{eqnarray}
k &=& \sqrt{k_z^2+k_t^2},\\
k_{\pm}&=&\frac{\sqrt{k_z^2+k_t^2} \pm k_z}{2}.
\end{eqnarray}      
    
\noindent We find directly the polar-coordinate representation of the radial-momentum operator in momentum space:

\begin{align}
\hat{N}_{k} = \frac{1}{2}\Big(k_t \frac{\partial}{\partial k_t}+\frac{i}{\sigma}\frac{\partial}{\partial k_{\phi}} - (k - k_z)\big(\frac{\partial}{\partial k_z} + \frac{1}{k} -\frac{w^2\Omega}{c} \big)\Big).\label{big} 
\end{align}  

This is the most exact form of the operator. Now, if we take the case of paraxial photons (which is in most cases the situation of interest), the momentum along the direction of propagation will be many orders of magnitude larger than the transverse momentum. So if we Taylor-series expand the square-root in $k_{-}$ and $k_{+}$ about $k_{z}$ we have

\begin{eqnarray}
k_{-}&=&\frac{1}{2}\left(\sqrt{k_{t}^{2}+k_{z}^{2}}-k_{z}\right),\nonumber\\
&=&\frac{1}{2}\left(k_{z}+\frac{k_{t}^{2}}{2k_{z}}-\frac{k_{t}^{4}}{8k_{z}^{3}}...-k_{z}\right),\nonumber\\
&\approx&\frac{k_{t}^{2}}{4k_{z}}.\\
k_{+}&=&\frac{1}{2}\left(\sqrt{k_{t}^{2}+k_{z}^{2}}+k_{z}\right),\nonumber\\
&=&\frac{1}{2}\left(k_{z}+\frac{k_{t}^{2}}{2k_{z}}-\frac{k_{t}^{4}}{8k_{z}^{3}}...+k_{z}\right),\nonumber\\
&\approx&k_{z}.
\end{eqnarray}

\noindent As well as $k=\sqrt{k_{t}^{2}+k_{z}^{2}}\approx k_{z}$. Now, again take Eq.(\ref{momwave}) \--- or use directly Eq.(\ref{big}) \--- and re-write it in terms of these more familiar momentum coordinates

\begin{eqnarray}
\psi^{\sigma}\approx e^{i\sigma mk_{\phi}}\delta\left(k_{z}-\frac{\Omega}{c}\right)\left(\frac{k_{t}^{2}}{2k_{z}}\right)^{n+\frac{|m|}{2}}e^{-\frac{w^{2}\Omega}{c}\frac{k_{t}^{2}}{2k_{z}}}k_{z}
\end{eqnarray} 

\noindent And making use of the delta function, we have up to constants,

\begin{eqnarray}
\psi^{\sigma}(k_{t},k_{z},k_{\phi})\propto e^{i\sigma mk_{\phi}}k_{t}^{2n+|m|}e^{-\frac{w^{2}}{2}k_{t}^{2}}.
\end{eqnarray} 

\noindent From this, unlike in the position-space paraxial-regime case, it is quite straightforward to find the operator which returns the value $n$. By simple inspection we can write

\begin{eqnarray}
\hat{N}'_{k}=\frac{1}{2}\left(k_{t}\frac{\partial}{\partial k_{t}}+\frac{i}{\sigma}\frac{\partial}{\partial k_{\phi}}+w^{2}k_{t}^{2}\right).\label{final}
\end{eqnarray}

\noindent Which has the most straightforward form. The first term is similar to the hyperbolic momentum operator, however it lacks the $i$ in front. This is non-trivial as without the $i$ it's action on the radial coordinate is given by 

\begin{eqnarray}
e^{i\gamma k_{t}\frac{\partial}{\partial k_{t}}/\hbar}\hat{r}e^{-i\gamma k_{t}\frac{\partial}{\partial k_{t}}/\hbar}=\hat{r}e^{i \gamma}.
\end{eqnarray}  

Where the radial position coordinate is defined in the momentum representation as

\begin{eqnarray}
\hat{r}\equiv -i\hbar \frac{\partial}{\partial k_{t}}.
\end{eqnarray}

Obviously this is problematic as the radial coordinate must remain a real number. Also, without the $i$ the the first term (by itself) is not in general hermitian. It's potentially interesting to note that if the action ($\gamma$) is a multiple of $\pi$ then it generates a real (periodic) scaling. Whether this can have some physical meaning is unknown to us as of yet.     

The second term is again the OAM operator, scaled by the helicity. The final term is the transverse laplacian in momentum coordinates scaled by the beam waist at origin. There are also a couple other advantages of writing the operator in this way. It is propagation invariant, that is, this is the operator for all time (as the momenta are conserved quantities). Also, the transverse momentum coordinate may be rescaled (as $k_{t}'=w\,k_{t}$) to make the operator independent of the beam waist \--- should this be desirable. 

We want to test is its hermiticity (in total). Does $\langle \hat{N}'_{k} \psi^{\sigma}|\psi^{\sigma}\rangle=\langle\psi^{\sigma}|\hat{N}'_{k} \psi^{\sigma}\rangle$? A straightforward computation shows that if we desire real eigenvalues, then $\hat{N}'_{k}$ can not, in general, be hermitian. It is hermitian instead only on a restricted class of wavefunctions where $\psi(k_{t},k_{\phi})=\Theta(k_{t})\Phi(k_{\phi})$, where $\Theta(k_{t})$ is a real function. Only in such beams can $\hat{N}_{k}$ be considered a valid observable. Some examples of such beams are the Bessel beams and the Laguerre-Gauss beams. Which is our main interest.

\section{Physical interpretation and technological prospects}  

Now that we have laid down the mathematical foundations for the radial index operator we can investigate ``meaning'' of the radial index. Though the RSV momentum-representation version of the operator, Eq.(\ref{final}), is accurate and simple, it's interpretation is more difficult to parse than the paraxial coordinate-representation operator Eq.(\ref{zdep}). This is indicative of the fact that the radial-index operator is \emph{not} representative of a general quantum observable like it's ``partner'' the OAM operator. Instead $n$ can be considered a proper observable in a restricted class of scenarios. More specifically for beams whose momentum-space wave-function is separable into radial and angular portions, and for which the radial part is always real. The radial-index operator also requires additional information to write down \--- specifically the beam-waist at origin, the propagation distance, and the orbital angular momentum.

This latter point has implications for future potential uses of the radial index as a tool in quantum technologies. Suppose we wish to measure the (unknown) radial index of a single photon. A common technique for doing this would be to shine the photon on a phase pattern which takes a photon of the radial index in question and turns it into the fundamental Gaussian mode, this photon then either couples \--- or does not couple \--- to a single mode fiber. This constitutes a projective measurement onto the chosen radial number. However, in order to choose the correct phase pattern, the OAM index, the beam waist, and the distance of propagation must be known. Note that this is not merely a problem with a particular experimental implementation of radial-index-sorting, but a general problem. If we examine the paraxial version of the operator \--- Eq.(\ref{zdep}) \--- we can see that it contains all three of the other parameters ($z$, $w_{o}$, and $l$) meaning that we must know these values a priori (or measure them simultaneously) in order to get a meaningful result from a radial-index measurement. It may also be possible to perform a measurement if the other parameters can be measured \emph{passively}, that is, non-destructively for the same photon (for example, OAM may be passively measured with a mode sorter \cite{sort}). 

To further illuminate this problem it should be noted that Laguerre-Gauss beams of different radial-index do not have zero overlap if they are at different distances in their propagation. Mathematically, that is 

\begin{eqnarray}
\int r\,dr\,\int d\phi \mathrm{LG}_{nl}^{*}(r,\phi,z)\mathrm{LG}_{n'l}(r,\phi,z')\neq 0.
\end{eqnarray}

\begin{figure}
\includegraphics[scale=0.41]{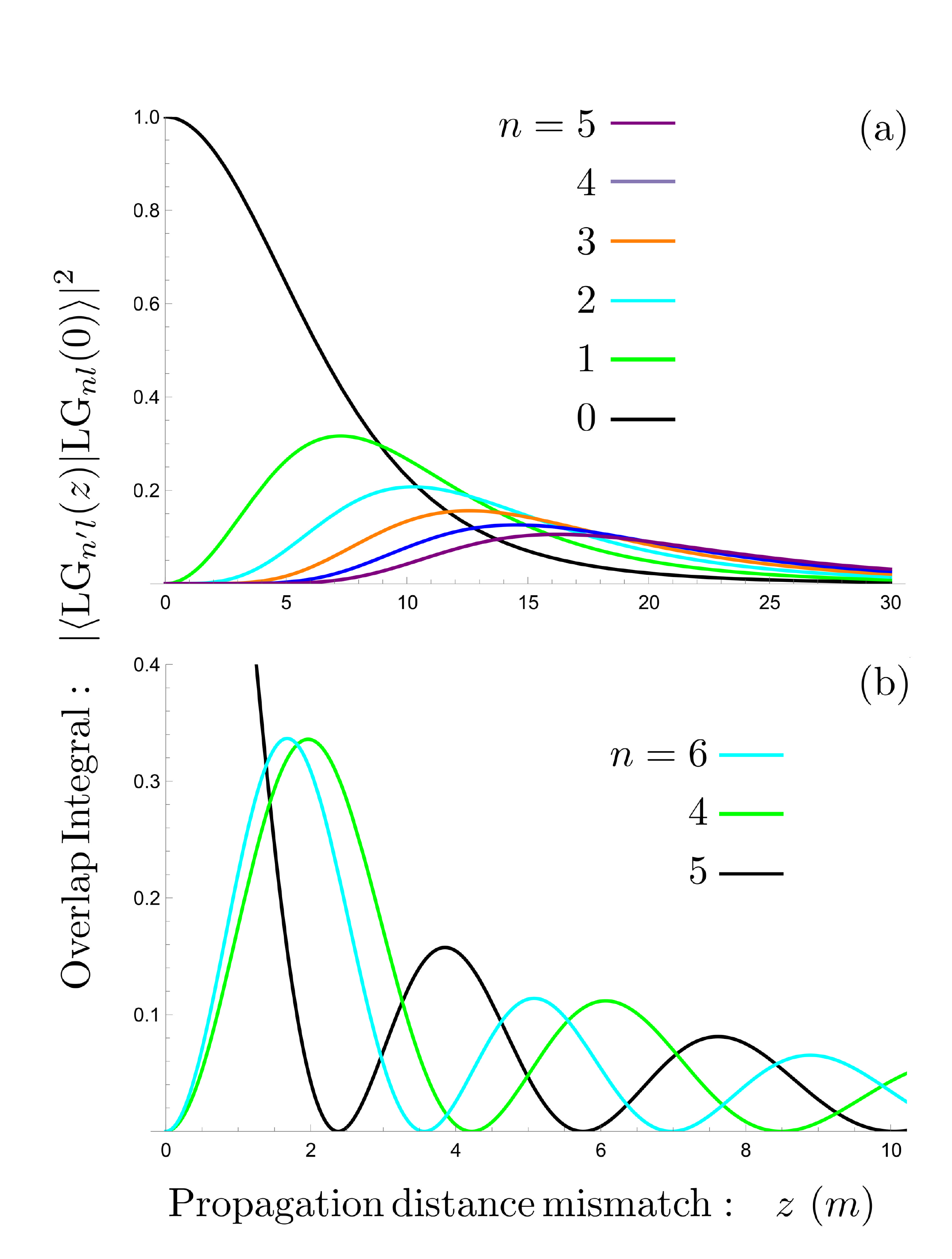}
\caption{Graphs of the overlap function of two Laguerre-Gauss modes with un-equal radial indices as a function of the mismatch in the distance the modes have propagated. In both graphs the black line represents the mode's overlap with itself. In (a) we show the $n=0$ mode's overlap with the five next higher indexed modes. It is clear from the graph that very quickly the depicted modes no longer add to one. This is due to the fact that as the beam propagates it acquires non-trivial overlaps with more and more modes. After only about 15 meters 20 modes are necessary to describe the beam to good accuracy (not shown). In (b) we show the overlap of an $n=5$ mode with a mode of higher index and a mode of lower index. Scaling is different than in (a) to highlight detail. The oscillatory behavior is due to the fact that after ``moving'' from one mode to an adjacent one, the photon may then ``move back'' as it continues to spread in the radial-index space.}
\label{overlap}
\end{figure}

In Figure \ref{overlap} we plot the overlap of a beam of a particular $n$ with several others for increasing propagation-distance mismatches. 

However if these problems can be overcome (which can be achieved by careful calibration) then indeed the radial-index can be used as a carrier of quantum information in realistic scenarios. This has been demonstrated recently by two separate experiments, as discussed earlier. 

These experiments show that the radial index can be a valuable resource in quantum experiments, but one needs to be careful to satisfy the subtle requirements in the measurements.        

The maxim ``One man's noise is another man's data'' may also find some application for the radial modes. To wildly speculate, it could be the case that the index's sensitivity to distance and beam-width mismatch could find some application, and that by (for example) measuring one $n$-mode's projection onto $n\pm 1$ some information about propagation distance or dilation could be obtained. 

Returning to the question of physical meaning, it is more illuminating to consider the paraxial version of the operator, Eq.(\ref{zdep}). Which we re-write below for the the convenience of the reader

\begin{eqnarray}
\hat{N}_{z}=-\frac{\hbar w_{z}^{2}}{8}\nabla^{2}_{t}-\frac{z}{kw_{o}^{2}}\widehat{P_{H}}-\frac{\hat{L}_{z}}{2}+\frac{\hbar}{2}\left(\frac{r^{2}}{w_{o}^{2}}-1\right).\label{re}
\end{eqnarray}

Where we have multiplied through by $\hbar$ and substituted the second term with the hyperbolic momentum operator \--- Eq.(\ref{hype}). There are four terms, three of which are related to fundamental properties of the beam which can be identified with quantities other than \--- and independent of \--- the radial index. That is to say that these three parameters influence \--- but are not influenced by \--- the radial index. This leaves a single term which represents the degree of freedom that the radial index truly represents. The first term in Eq.(\ref{re}) is the transverse Laplacian scaled by the $z$-dependent beam waist and a numerical constant. In the free-space paraxial regime the transverse Laplacian alone is the Hamiltonian (and, of course, a constant of motion). The third term is the OAM. The final term produces \--- up to constants \--- the second moment of the radial position at $z=0$, which is clearly related to the transverse spatial variance, i.e. the spatial confinement at origin. These are three independent parameters of the beam (and of individual photons which occupy these modes). 

It is the second term, however, which is most important for understanding the physical meaning of the radial index. This is the hyperbolic momentum operator which generates dilations. As discussed previously it is a mathematically well-formed kind of ``radial momentum operator''. It is however \emph{not} a constant of motion (as is demonstrated in Fig.\ref{phz}), and thus does not have one of the quantities we would traditionally expect from a momenta. Nonetheless, if the propagation distance is accounted for in the operator itself (as is the case in $\hat{N}_{z}$), then it is sensible to speak of it as an intrinsic property; here we see why the operator must be $z$-dependent in the coordinate representation and becomes somewhat unmanageable in the momentum representation. In other words, the radial index is best understood in the coordinate picture, as in that representation the fact that it is not conserved under time-evolution is easy to account for, however the momentum representation deals best with quantities that are preserved for all time.  

The radially transverse momentum of a photon must be dependent on the total transverse momentum, as well as the orbital angular momentum (as the orbital part of the total transverse momentum needs to be compensated for); and on the degree of spatial confinement of the beam (because greater confinement must lead to higher divergence and greater radial momentum). However this does not account for intrinsic freedom of the hyperbolic momentum. The radial index can then be seen as representing the promotion of the radial-like, hyperbolic momentum \emph{itself} to a quantum observable. We therefore believe that the radial index could best be described as the ``intrinsic hyperbolic momentum charge''.  

We might have initially expected the radial index to be related to the radial momentum, as the azimuthal index is related to a rotational momentum, completing the set of two momenta necessary to move about in the transverse 2-D plane. We have found that, while $\hat{N}_{z}$ does not correspond to a full radial momentum operator (as no such well-formed operator exists), the hyperbolic momentum operator \--- which is up to scalings and other observables the operator we find \--- has a direct connection to the momentum along the radial coordinate. 

It has been shown \cite{lof2} that the radial index is conserved in the process of spontaneous parametric down-conversion. More specifically, in the ideal case, the two daughter photons will have the same radial index as each other. In the non-ideal case \--- finite crystal length and width \--- there is a spread in the radial index of the down-converted photons. As we have seen in the research in this manuscript, the radial index can only be said to be well-defined given a precise propagation distance, OAM, and beam-width; therefore we gain insight as to why the effect of crystal length (and thus uncertainty in the point of origin of the down-conversion; and walk-off) should ``blur'' the resultant radial indices. This is in addition to a similar effect from a limited transverse spatial extent of the crystal \--- investigated in detail in Ref.\cite{trans}   

Since this paper and the previous work of ourselves and others have shown that both the orbital angular momentum and the intrinsic hyperbolic momentum charge are well-formed quantum numbers with a physical meaning (in the proper circumstances), it follows that that the mode indices of beams with different symmetries (non-circular cylindrical, e.g. elliptic or Cartesian) are also well-formed quantum numbers. These other mode indices are just algebraic combinations of $l$ and $n$. Therefore, it should follow that these other numbers should have interesting physical meanings as well. For the case of paraxial photons with elliptic symmetry \--- the so-called Ince-Gauss modes \cite{mex} \--- it has been shown that their mode indices can be be entangled \cite{mi}. Also, that their OAM properties are non-trivial and show interesting complexity \cite{bi}. It may be the case that these effects could be better understood by combining the concepts of orbital and hyperbolic momentum.  

There are still some unanswered questions \--- most significantly \--- it is not clear should be considered the conjugate variable. Also, we do not yet understand why the ``intrinsic hyperbolic momentum charge'' (radial index) should take discreet values. Furthermore it is unknown why the hyperbolic momentum should be tied to the number of rings in the transverse spatial pattern of the beam. 

We would also like to briefly sketch some potential technological applications and concerns tied to the radial index. As we have mentioned already it has been shown experimentally that the radial index is indeed a valid quantum number \cite{ke}, and could be used for quantum communication. However the fact that different $n$-modes can have non-zero overlap if there is mismatch in the beam waist $w_{o}$, or propagation distance $z$, adds some additional complication, especially in realistic scenarios where complete control or knowledge of the distances and amount of focusing involved may not be available. However, if we take the $z=0$ form of the radial-index operator, Eq.(\ref{N0}), we find that this is exactly the Hamiltonian of a graded-index fiber (GRIN) \cite{gabby}. GRINs are commercially available fibers that have a linear radial variation of their refractive indices. It is possible that a hyperbolic-momentum carrying photon could be matched with standard optics to such a fiber (i.e. beam waist matched to the correct steepness of index variation) then the radial index would be intrinsically preserved under propagation in the fiber. This would allow for quantum or classical communication multiplexing using the radial degree of freedom in a widely-available kind of fiber. This is an exciting prospect that we believe merits further investigation.               

\section{Summary and Conclusions}

In this manuscript we have developed the differential-operator formalism for the radial index of Laguerre-Gauss modes in both the paraxial, coordinate representation and in the exact, momentum-space representation. We identified the various parts of these operators with certain physical parameters, most of which are tied to beam characteristics which are not influenced by the radial index but are influential on it. Put another way, for each value of these parameters (beam waist, orbital angular momentum, propagation distance) there is a different representation of the radial-index operator. There is one remaining part of the operator which is not tied to other properties of the photon which we have identified: the hyperbolic momentum which generates dilations. It is this term which corresponds to the true degree of freedom the radial index represents. We thus call the the radial index the ``intrinsic hyperbolic momentum charge''. We have shown that the radial index is not tied exclusively to the transverse spatial profile of the beam but also has a physical meaning. We hope that this opens up a new area of investigation and inspires some new potential technological prospects.

We briefly outlined one such potential application, which is that the radial index may be naturally preserved under propagation in a graded-index fiber. This may have application to fiber-optic multiplexing in the classical and quantum domains. 

In conclusion we conjecture that, despite the resources devoted to the study of the orbital angular momentum degree of freedom, the \emph{radial} degree of freedom is vastly richer due to its mathematical complexity, and the plethora of questions that remain unanswered. 

\section*{Acknowledgments}

\noindent This work was supported by the Austrian Academy of Sciences (ÖAW), the European Research Council (SIQS Grant No. 600645 EU-FP7-ICT), and the Austrian Science Fund (FWF) with SFB F40 (FOQUS). W.N.P. also received funding from ERC Advanced Grant QIT4QAD, and the French National Research Agency through the project COMB (grant number ANR-13-BS04-0014). Discussions with Sven Ramelow, Anton Zeilinger, Jason Twamley, Robert Fickler, Gabriel Molina-Terriza, Iwo Bialynicki-Birula, and Ebrahim Karimi were invaluable.

\section*{Appendix: Derivation of the photon wave-function}

This section is for the most part a summary of the formalism collected or derived in Ref.\cite{beam}. Please note however that some definitions and notation will differ.

In the RS vector formalism, Maxwell's equations in free space reduce to

\begin{eqnarray}
\frac{\partial}{\partial t}\vec{F}\left(\vec{x},t\right)&=&-ic\nabla\times\vec{F}\left(\vec{x},t\right),\label{max1}\\
\nabla\cdot\vec{F}\left(\vec{x},t\right)&=&0.
\end{eqnarray}

\noindent The RS vector can also be expressed as a complex vector field $\vec{Z}$:

\begin{eqnarray}
\vec{F}\left(\vec{x},t\right)=\nabla\times\left(\frac{i}{c}\vec{Z}\left(\vec{x},t\right)+\nabla\times\vec{Z}\left(\vec{x},t\right)\right).
\end{eqnarray}

\noindent The function $\vec{Z}$ satisfies the wave equation

\begin{eqnarray}
\left(\frac{1}{c^{2}}\frac{\partial}{\partial t}-\nabla^{2}\right)\vec{Z}\left(\vec{x},t\right)=0.\label{wave}
\end{eqnarray}

\noindent Where $\nabla^{2}$ is the full laplacian. The potential $\vec{Z}$ can, itself, be recast in a more tractable form: $\vec{Z}=(e_{1},e_{2},e_{3})\chi (\vec{x},t)$, where $\chi$ is a field and the $e$'s may be chosen to suit the situation at hand (i.e. the desired symmetries of the system to be studied), for beams propagating in a straight line $\vec{Z}=(0,0,1)\chi (\vec{x},t)$, is common. 

All solutions to the wave equation \--- Eq.(\ref{wave}) \--- may be expressed as a superposition of plane waves, so 

\begin{eqnarray}
\chi (\vec{x},t)&=&\int d\vec{k}\,N(\vec{\,k\,})\label{scal1}\\
& &\times\left(\psi^{+}(\,\vec{k}\,)\,e^{-i\omega_{k}t+i\vec{k}\cdot\vec{x}}+\psi^{-}(\,\vec{k}\,)\,e^{i\omega_{k}t-i\vec{x}\cdot\vec{r}}\right).\nonumber
\end{eqnarray}

\noindent Where $N(\vec{k})$ represents a normalization factor. Then the RS vector can be written as

\begin{eqnarray}
\vec{F}\left(\vec{x},t\right)&=&\int d\vec{k}\,\,\vec{e}(\vec{k})\label{F}\\
& &\times\left(\psi^{+}(\,\vec{k}\,)\,e^{-i\omega_{k}t+i\vec{k}\cdot\vec{x}}+\psi^{-}(\,\vec{k}\,)\,e^{i\omega_{k}t-i\vec{x}\cdot\vec{r}}\right).\nonumber
\end{eqnarray}

\noindent Where $\vec{e}$ is a $k$-dependent complex polarization vector which is determined by the choice of the vector part of the potential $\vec{Z}$ and includes the normalization factor. This represents a gauge freedom. Note that the polarization vector factors completely from the rest of the expression. It can thus be considered a part of the transformation from momentum space to position space. This is a special feature of the RS formulation. The weight functions $\psi^{-}(\,\vec{k}\,)$ and $\psi^{+}(\,\vec{k}\,)$ are the positive and negative frequency components of the RS vector in momentum-space, the choice of which completely define the physical degrees of freedom of the EM field in momentum space. 

In our treatment we wish for a fully well-formed construction, so we will have as our objective to ``end up'' in momentum-space, since it has been long-known that the definition of the photon wave-function in position space is problematic. This is due to the fact that a position operator does not exist for photons (eigenstates of momentum for photons \--- plane waves \--- though non-physical can be mathematically tractable, but the idea of a position eigenstate for photons is non-nonsensical even at the conceptual level as photons are inherently de-localized). A \emph{probabilistic} position representation can generated for photons but we will not take this route. For more information on this and connected topics please see Refs.\cite{BB1}\cite{BB2}.

However, for reasons of clarity we will start in position space and then switch to momentum. The time-dependent Schr\"{o}dinger equation is

\begin{eqnarray}
i\hbar\,\frac{\partial}{\partial t}\Psi\left(\vec{x},t\right)=\widehat{H}\,\Psi\left(\vec{x},t\right).
\end{eqnarray}      

\noindent Where $\widehat{H}$ is the Hamiltonian. One immediately recognizes the similarity between this and Eq.(\ref{max1}), the first Maxwell's equation for the RS vector. We can thus equate the Hamiltonian to the appropriate part of that equation $\widehat{H}\,\Psi=-ic\nabla\times\,\Psi$ with the RS vector becoming the photonic wave-function. 

In order to proceed from here we will rewrite the curl operator. Take a vector $\hat{\vec{s}}$, the components of which are the spin-1 matrices

\begin{eqnarray}
\hat{s}_{x}=\left[ 
\begin{array}{ccc}
0 & 0 & 0 \\
0 & 0 & -i \\
0 & i & 0 
\end{array} 
\right],
\hat{s}_{y}=\left[ 
\begin{array}{ccc}
0 & 0 & i \\
0 & 0 & 0 \\
-i & 0 & 0 
\end{array} 
\right],
\hat{s}_{z}=\left[ 
\begin{array}{ccc}
0 & -i & 0 \\
i & 0 & 0 \\
0 & 0 & 0 
\end{array} 
\right].\nonumber
\end{eqnarray}   

\noindent Written in index summation notation the tensor $\hat{\vec{s}}=s_{ijk}$ is both equivalent to the Levi-Civita symbol $\epsilon_{ijk}$ times a factor of $-i$, and the spin-1 matrices in quantum mechanics (when these matrices are acting on the Cartesian vector components of the wave-function and not the eigenstates of $\hat{s}_{z}$). Since it is the case that $\vec{a}\times\vec{b}=\epsilon_{ijk}a_{j}b_{k}$ and $\vec{a}\cdot\vec{b}=a_{i}b_{i}$, it is also the case that

\begin{eqnarray}
\nabla\times\,\Psi=-i\hat{\vec{s}}\cdot\nabla\,\Psi.\label{ham}
\end{eqnarray}  

\noindent Where $\hat{p}=-i\hbar\nabla$ is the very well-known momentum operator. Slightly less well-known is the ``helicity'' operator in optics and particle physics, which is the sign of the projection of the angular momentum on the momentum 

\begin{eqnarray}
\hat{\Lambda}&=&\mathrm{Sgn}\left[\widehat{M}\cdot\hat{p}\right],\\
&=&\mathrm{Sgn}\left[\left(-i\hbar\vec{r}\times\nabla+\hbar\hat{\vec{s}}\right)\cdot\left(-i\hbar\nabla\right)\right],\\
&=&\frac{1}{p_{T}}\hat{\vec{s}}\cdot\hat{p}.
\end{eqnarray}  

\noindent Where the momentum and angular momentum operators are substituted for their definitions on the second line (the later is composed of an orbital and spin part), $p_{T}=\sqrt{p_{x}^{2}+p_{y}^{2}+p_{z}^{2}}$, and the simplification on the last line is due to the fact that $(\vec{r}\times\nabla)\cdot\nabla=0$. So we can see directly by comparing the above to Eq.(\ref{ham}) that the helicity operator (up to constants) is the Hamiltonian. The helicity operator generates the duality rotation and is associated with that symmetry in free-space. Other than the fact that it serves as the Hamiltonian for a Schr\"{o}dinger equation with the RS vectors as wave-functions, the properties of $\hat{\Lambda}$ are not strictly relevant to our work here, nonetheless the interested reader is re-directed to Refs.\cite{H1}\cite{H2}.   

It will be convenient to find the eigenfunctions of the helicity (and thus the energy eigenstates) in a beam-like basis. We return to Eq.(\ref{F}) and utilize the following expansion in cylindrical coordinates

\begin{eqnarray}
e^{i\vec{k}\cdot\vec{x}}=e^{ik_{z}z}\sum_{m=-\infty}^{\infty}i^{m}e^{im(\phi-k_{\phi})}J_{m}(k_{t}r).
\end{eqnarray} 

\noindent Where $k_{\phi}$ and $k_{t}$ are the polar and radial coordinates in momentum space, respectively. Note that the radial coordinate in momentum space is also the transverse momentum. The functions $J_{m}$ are Bessel functions of the first kind. This affects a change of basis from plane waves to Bessel waves, which also forms a complete set. Now Eq.(\ref{scal1}) becomes

\begin{eqnarray}
\chi (\vec{x},t)=\sum_{m,\sigma}\int d\vec{k}\,\,\chi^{\sigma}_{m}(\,\vec{k}\,)\,\psi^{\sigma}(\,\vec{k}\,).\label{xmo}
\end{eqnarray}

\noindent Where $\sigma$ can take values $\pm 1$ and the $\chi^{\pm}_{m}$'s are defined as 

\begin{eqnarray}
\chi^{\pm}_{m}(\vec{x},\vec{k}\,)=\frac{(\pm i)^{m}}{k\,k_{t}\sqrt{2}}e^{\pm i (\omega_{k}t-k_{z}z-m(\phi-k_{\phi})}J_{m}(k_{t}r).\label{xm}
\end{eqnarray}

\noindent We then obtain via choosing the beam-like form of the potential ($\vec{Z}=(1,0,0)\chi$) and by direct calculation for a particular $m$ and $\vec{k}$.  

\begin{eqnarray}
\vec{F}^{\sigma}_{k_{t}k_{z}m}=\left[
\begin{array}{c}
F_{r} \\
F_{\phi} \\
F_{z} 
\end{array} 
\right]
&=&\frac{(i\sigma)^{m}}{k\sqrt{2}}e^{-i\sigma(\omega_{k}t-k_{z}z-m\phi)}\label{Fb}\\
& &\times\left[
\begin{array}{c}
i\sigma k_{z}\frac{\partial}{\partial(k_{t}r)}+i\frac{km}{k_{t}r} \\
-\sigma k\frac{\partial}{\partial(k_{t}r)}-\frac{k_{z}m}{k_{t}r} \\
k_{t}
\end{array} 
\right]J_{m}(k_{t}r).\nonumber
\end{eqnarray}

\noindent The number $\sigma$ represents whether the photon is right or left circularly polarized. It is the case that $\hat{\Lambda}\vec{F}^{\sigma}_{k_{t}k_{z}m}=\sigma\vec{F}^{\sigma}_{k_{t}k_{z}m}$. The wave-functions in Eq.(\ref{Fb}) form a basis of solutions to the photonic Schr\"{o}dinger's equation. It is important to note that, since the helicity takes the place of the Hamiltonian, there are solutions to the Schr\"{o}dinger's equation which at first glance might seem to have negative energy. These solutions (those of left-circularly-polarized photons) do not represent anti-particles as photons have no anti-particles. They should be interpreted merely as opposite helicity photons. These beam-like solutions are in fact the Bessel beams, which have come to be objects of interest due to their diffraction-free properties \cite{diff}. However, much like plane-waves, these beams are not physical as they turn out to have infinite energy. Typically these beams are made physical without breaking the assumption of monochormicity by convolving them with a Gaussian function (the Bessel-Gauss beams). However these beams are no longer exact solutions to the Maxwell's equations. However by dropping the assumption of a monochromatic beam (also usually implicit in the paraxial wave-equation) we can use the Bessel-beam basis defined in Eq.(\ref{Fb}) and continue to have exact solutions to the photon wave-function which are valid under all circumstances.


\begin{thebibliography}{10}

\bibitem{allen}``Orbital angular momentum of light and the transformation of Laguerre-Gaussian laser modes'' L. Allen, M.W. Beijersbergen, R.J.C. Spreeuw, and J.P. Woerdman, Phys. Rev. A \textbf{45}, 8185 (1992).\\

\bibitem{rev}``Advances in optical angular momentum'' S. Franke-Arnold, L. Allen, and M. Padgett, Laser and Photonics Reviews \textbf{2}, 299 (2008).\\

\bibitem{plick}``The Forgotten Quantum Number: A short note on the radial modes of Laguerre-Gauss beams'' W.N. Plick, R. Lapkiewicz, S. Ramelow, A. Zeilinger,  arXiv:1306.6517 (2013).\\

\bibitem{k1}``Radial coherent and intelligent states of paraxial wave equation'' E. Karimi, and E. Santamato, Opt. Lett. \textbf{37}, 2484 (2012).\\

\bibitem{k2}``Lost and found: the radial quantum number of Laguerre-Gauss modes'' E. Karimi, R. W. Boyd, P. de la Hoz, H. de Guise, J. Rehacek, Z. Hradil, A. Aiello, G. Leuchs, and L. L. Sanchez-Soto, Phys. Rev. A \textbf{89}, 063813 (2014).\\ 

\bibitem{ke}``Exploring the quantum nature of the radial degree of freedom of a photon via Hong-Ou-Mandel interference'' E. Karimi, D. Giovannini, E. Bolduc, N. Bent, F.M. Miatto, M.J. Padgett, and R.W. Boyd, Phys. Rev. A \textbf{89}, 013829 (2014).\\

\bibitem{BB1}``Photon wave function'' I. Bialynicki-Birula, Prog. Opt. \textbf{36}, 245 (1996).\\  

\bibitem{BB2}``The role of the Riemann-Silberstein vector in classical and quantum theories of electromagnetism'' I. Bialynicki-Birula, and Z. Bialynicki-Birula, J. Phys. A: Math. Theor. \textbf{46}, 053001 (2013).\\

\bibitem{beam}``Beams of electromagnetic radiation carrying angular momentum: The Riemann-Silberstein vector and the classical-quantum correspondence'' I. Bialynicki-Birula, and Z. Bialynicki-Birula, Opt. Comm. \textbf{264}, 342 (2004).\\

\bibitem{calvo}``Quantum field theory of photons with orbital angular momentum'' G. F. Calvo, A. Pic´on, and E. Bagan, Phys. Rev. A 73, 013805 (2006).\\

\bibitem{abs}``Direct Observation of Transfer of Angular Momentum to Absorptive Particles from a Laser Beam with a Phase Singularity'' H. He, M.E.J. Friese, N.R. Heckenberg, and H. Rubinsztein-Dunlop, Phys. Rev. Lett. \textbf{75}, 826 (1995).\\

\bibitem{first}``Generalized Momentum Operators in Quantum Mechanics'' P.D. Robinson, and J.O. Hirschfelder, J. Math. Phys. \textbf{4}, 338 (1963).\\

\bibitem{T1}``Quantum distribution functions for radial observables'' J. Twamley, J. Phys. A: Math. Gen. \textbf{31}, 4811 (1998).\\

\bibitem{T2}``The Quantum Mellin transform'' J. Twamley, and G.J. Milburn, New J. Phys. \textbf{8}, 328 (2006).\\

\bibitem{lof}``Full-Field Quantum Correlations of Spatially Entangled Photons'' V.D. Salakhutdinov, E.R. Eliel, and W. L\"offler
Phys. Rev. Lett. \textbf{108}, 173604 (2012).\\

\bibitem{mario}``Generation and confirmation of a (100 X 100)-dimensional entangled quantum system'' M. Krenn, M. Huber, R. Fickler, R. Lapkiewicz, S. Ramelow, and A. Zeilinger, Pro. Net. Aca. Sci. \textbf{111}, 6243 (2014).\\

\bibitem{walsh}``Walsh modes and radial quantum correlations of spatially entangled photons'' D. Geelen and W. L\"{o}ffler, Opt. Lett. \textbf{38}, 4108 (2013).\\

\bibitem{LGBB}``Laguerre-Gauss beams versus Bessel beams showdown: peer comparison'' J. Mendoza-Hern\'{a}ndez, M.L. Arroyo-Carrasco, M.D. Iturbe-Castillo, and S. Ch\'{a}vez-Cerda, Opt. Lett. \textbf{40}, 3739 (2015).\\

\bibitem{sort}``Efficient Sorting of Orbital Angular Momentum States of Light'' G.C.G. Berkhout, M.P.J. Lavery, J. Courtial, M.W. Beijersbergen, and M.J. Padgett, Phys. Rev. Lett. \textbf{105}, 153601 (2010).\\

\bibitem{trans}``Full characterization of the quantum spiral bandwidth of entangled biphotons'' F.M. Miatto, A.M. Yao, and S.M. Barnett, Phys. Rev. A \textbf{83}, 033816 (2011).\\

\bibitem{lof2}``Full-Field Quantum Correlations of Spatially Entangled Photons'' V. D. Salakhutdinov, E. R. Eliel, and W. Löffler, Phys. Rev. Lett. \textbf{108}, 173604 (2012).\\ 

\bibitem{mi}``Entangled singularity patterns of photons in Ince-Gauss modes'' M. Krenn, R. Fickler, M. Huber, R. Lapkiewicz, W.N. Plick, S. Ramelow, and A. Zeilinger, Phys. Rev. A \textbf{87}, 012326 (2013).\\

\bibitem{bi}``Quantum orbital angular momentum of elliptically symmetric light'' W.N. Plick, M. Krenn, R. Fickler, S. Ramelow, and A. Zeilinger, Phys. Rev. A \textbf{87}, 033806 (2013).\\

\bibitem{gabby}``Vortex revivals with trapped light'' G. Molina-Terriza, L. Torner, E.M. Wright, J.J. Garc\'{i}a-Ripoll and V.M. P\'{e}rez-Garc\'{i}a, Opt. Lett. 26, 1601 (2001).\\

\bibitem{mex}``Ince–Gaussian beams'' M. A. Bandres and J. C. Guti´errez-Vega, Opt. Lett. \textbf{29}, 144 (2004).\\

\bibitem{H1}``Optical helicity, optical spin and related quantities in electromagnetic theory'' R.P. Cameron, S.M. Barnett, and A.M. Yao, New J. Phys. \textbf{14}, 053050 (2012).\\

\bibitem{H2}``Dual electromagnetism: helicity, spin, momentum and angular momentum'' K.Y. Bliokh, A.Y. Bekshaev, and F. Nori, New J. Phys. \textbf{15}, 033026 (2013).\\

\bibitem{diff}``Bessel beams: diffraction in a new light'' D. McGloin, and K. Dholakia, Cont. Phys. \textbf{46}, 15 (2005).

\end{thebibliography}
\end{document}